# Graphene-Enhanced Thermal Interface Materials for Thermal Management of Photovoltaic Solar Cells

## M. Saadah, D. Gamalath, E. Hernadez and A.A. Balandin

Nano-Device Laboratory (NDL) and Phonon Optimized Engineered Materials (POEM) Center
Department of Electrical and Computer Engineering
University of California – Riverside
Riverside, California 92521 USA

## ABSTRACT

The increase in the temperature of photovoltaic (PV) solar cells affects negatively their power conversion efficiency and decreases their lifetime. The negative effects are particularly pronounced in concentrator solar cells. Therefore, it is crucial to limit the PV cell temperature by effectively removing the excess heat. Conventional thermal phase change materials (PCMs) and thermal interface materials (TIMs) do not possess the thermal conductivity values sufficient for thermal management of the next generation of PV cells. In this paper, we report the results of investigation of the increased efficiency of PV cells with the use of graphene-enhanced TIMs. Graphene reveals the highest values of the intrinsic thermal conductivity. It was also shown that the thermal conductivity of composites can be increased via utilization of graphene fillers. We prepared TIMs with up to 6% of graphene designed specifically for PV cell application. The solar cells were tested using the solar simulation module. It was found that the drop in the output voltage of the solar panel under two-sun concentrated illumination can be reduced from 19% to 6% when graphene-enhanced TIMs are used. The proposed method can recover up to 75% of the power loss in solar cells.

## I. INTRODUCTION

The importance of renewable energy cannot be overstated. The photovoltaic market has been increasing gradually over the past 10 years. The most widely used solar cells are the crystalline silicon accounting for almost 90% of the market [1-3]. For most commercial solar cells, about 15% of the incident solar energy is converted into useful power, while the rest is converted into heat inside the solar cell. Since heat has such a substantial negative effects on solar cells, several research attempts has been made to control it [4-14]. A popular cooling method is achieve by the use of heat spreaders or heatsinks. However, the microscopic imperfections between the two surfaces of the solar panel and heatsink gives air pockets which raises the thermal resistance between the two surfaces. The goal of this research is to fill these air gaps with a graphene-enhanced highly conducive thermal interface material (TIM) and thermal phase change materials (PCMs).

Two important parameter that can be experimentally measured in solar cells are open circuit voltage ($V_{OC}$) and short circuit current ($I_{SC}$). The open circuit voltage is the cell's maximum output voltage when the circuit is open and no current is flowing, while the short circuit current is the maximum current flowing when the cell is shorted. The efficiency of solar cell can be obtained using these two values. The change in open circuit voltage ($V_{OC}$) with temperature is given by [3]:

$$\frac{dV_{OC}}{dT} = \left(\frac{V_{OC}}{T}\right) + \frac{kT}{q}\left(\frac{1}{I_{SC}}\frac{dI_{SC}}{dT} - \frac{1}{I_0}\frac{dI_0}{dT}\right)$$ (1)

Where, $q$ is electron charge, $T$ is the temperature, $q$ is electron charge, and $I_0$ is the reverse saturation current. Another important parameter is the fill factor ($FF$). The fill factor correspond to the maximum power generated by a solar cell ($P_{mp}$) and is given by [15]:

$$FF = \frac{P_{mp}}{I_{SC}V_{OC}}$$ (2)

Most crystalline silicon solar cells have a fill factor of 0.7 - 0.85 [16]. This value is important to calculate the solar cell's efficiency ($\eta$):

$$\eta = \frac{P_{mp}}{P_{in}} = \frac{FF.I_{SC}V_{OC}}{P_{in}}$$ (3)

Where, $P_{in}$ is the incident solar power onto the PV solar cell.

## II. EXPERIMENT PROCEDURES

Graphene has proved itself as efficient thermal filler material for electronic and battery cooling applications [17-25]. In this work we extended the use of graphene as filler for TIMs and PCMs for thermal management of solar cells. A small percentage of graphene, about 1-6%, is mixed with a commercial TIM to increase its thermal conductivity. The weight percentage of mixed graphene is given by:

$$wt\% = \frac{W_g}{W_T + W_g}$$ (4)

Here, $W_T$ and $W_g$ are the weight of TIM and graphene respectively. The compound mixture is then applied between the solar panel and the heatsink to replace the microscopic air pockets caused by surface imperfections with far better heat conducting material. The pressure between the solar panel and the heatsink is maintained at 100 kPa during the experiment. The goal is to reduce the thermal resistance between the two surfaces which is govern by the following formula:

$$R_{TIM} = \frac{BLT}{K_{TIM}} + R_{C1} + R_{C2}$$ (5)

Where, $R_{TIM}$ is the thermal resistance, BLT is the bond-line thickness between the two surfaces, and $R_{C1}$, $R_{C2}$ are the contact resistance for the two surfaces as shown in Figure 1.

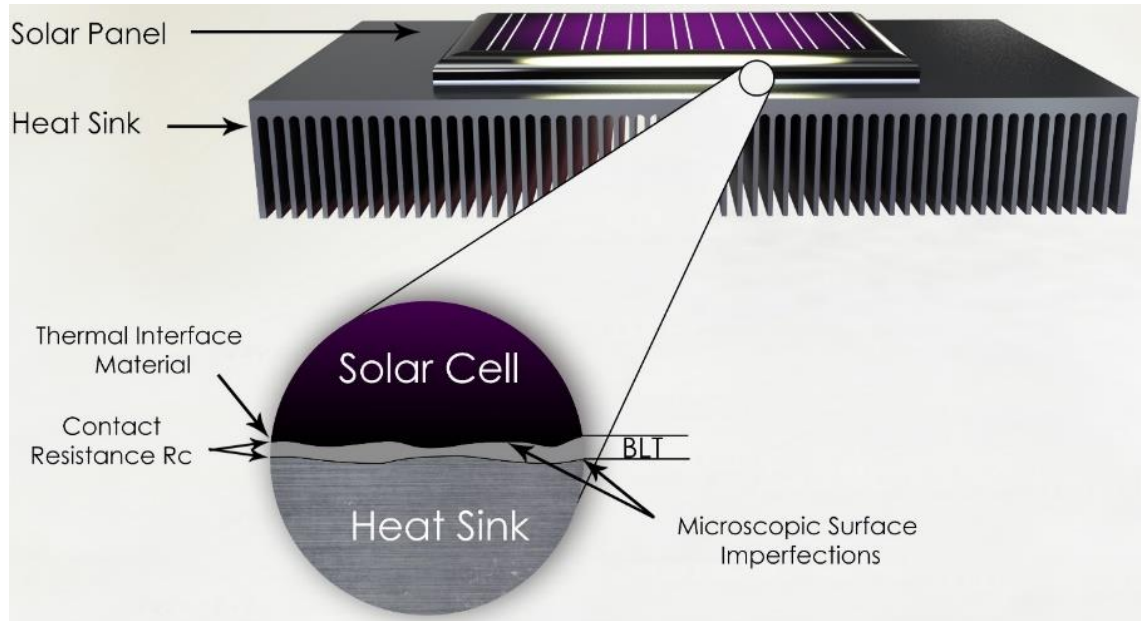

Figure 1: Illustration of operation of thermal interface material (TIM).

The solar panel is then placed under a high grade commercial solar simulator as shown in Figure 2. The solar simulator utilizes a Xenon lamp that produce a 6"x6" size beam that correspond to earth's 5800K blackbody spectrum. The uniformity of the illumination is insured by a highly regulated power supply. To concentrate the illumination, a convex lens is placed between the solar simulator and the solar cell. The illumination beam is concentrated between 1-5 suns when the solar cell is lowered/raised with a lab jack.

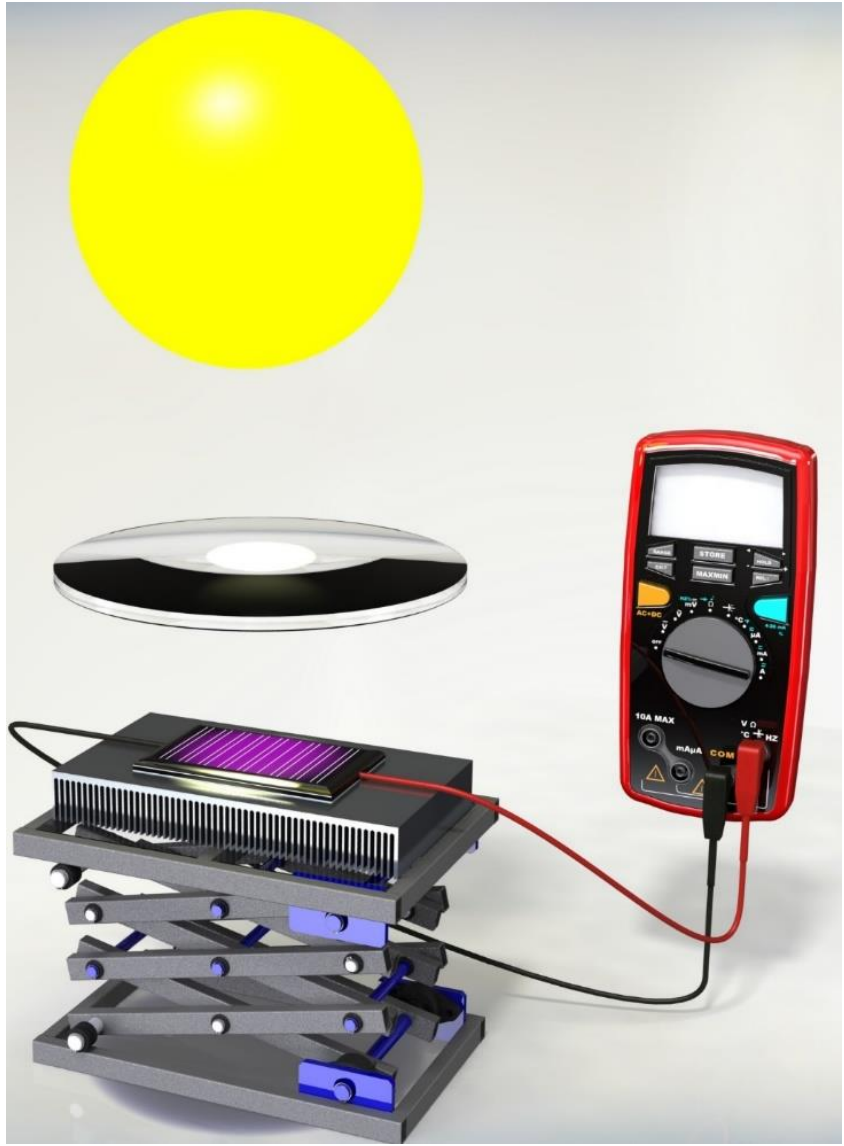

Figure 2: Illustration of the solar simulation setup used in this research.

The temperature of the cell, open circuit voltage ($V_{OC}$), and short circuit current ($I_{SC}$) are then measured and logged every second to observe the change due to the rise in cell temperature.

### III. RESULTS AND DISCUSION

The heat effects on a solar cell are evidently demonstrated in Figure 3. Once the solar cell is subjected to the sun beam, the cell temperature starts to rise gradually from 25 °C to 55 °C. This rise in cell temperature causes the open circuit voltage ($V_{OC}$) to drop gradually to about 12%. This translate to about 0.4% drop in efficiency per 1 °C increase in cell temperature. The solar simulator is then shut off and on several times to observe the changes in voltage due to cell heating up and cooling down. The effect of cell's temperature rise on the short circuit current ($I_{SC}$) was minimal. The current only changes less than 1% for the entire 30 °C change in cell temperature, which is in line with the theory and prior experimental literature [3, 15, 16, 18].

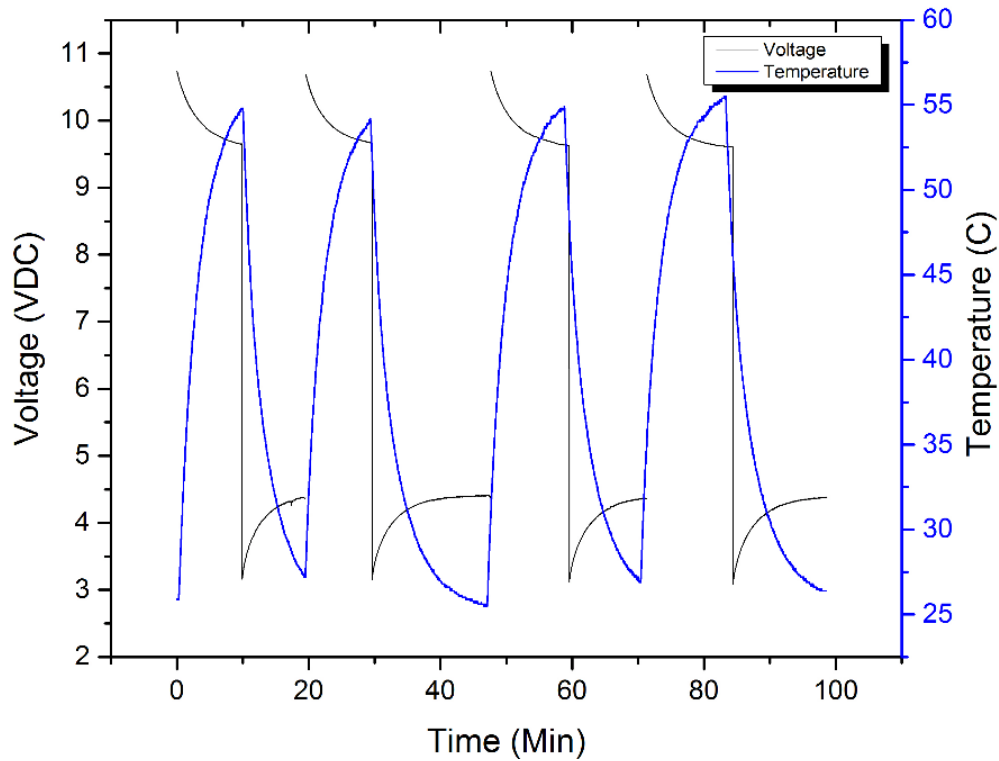

Figure 3: Temperature effects on open circuit voltage ($V_{OC}$).

The output voltage is then measured using a heatsink with and without TIM between its surface and the solar cell. Figure 4 shows that the drop in $V_{OC}$ is 3% with the 6 wt% graphene TIM compared to 12% drop without using any kind of heat removal. This cooling technique allows the recovery of about 75% of the power lost due to cell temperature rise.

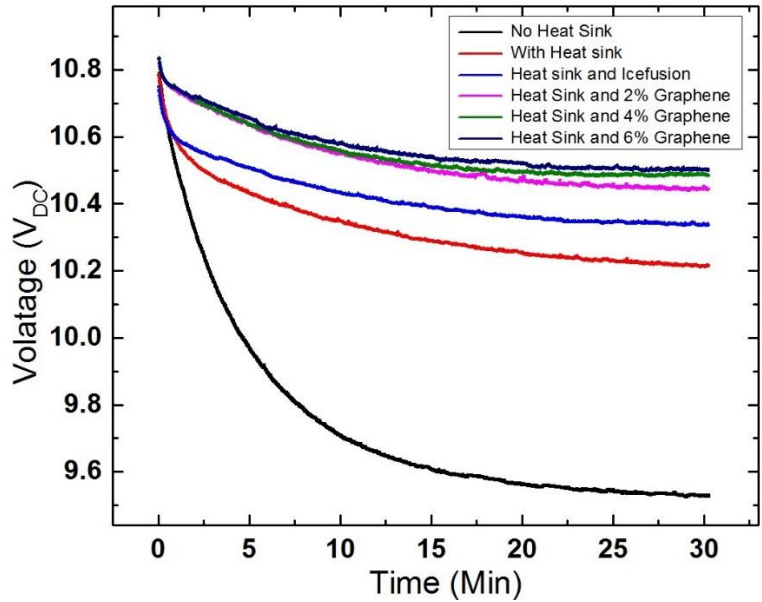

Figure 4: Open circuit voltage ($V_{OC}$) measured with and without cooling at 1 sun concentration.

Another smaller solar cell is tested under a concentrated illumination of 2 and 5 suns. The cell temperature was found to increase more with the concentration of solar light as shown in figure 5. The open circuit voltage raises slightly to about 6% when using a 5 sun concentration, figure 6, while the short circuit current raises substantially to about three folds compared to 1 sun intensity. Figure 7 and 8 shows the change in the open circuit voltage when using 2 and 5 suns respectively with and without cooling.

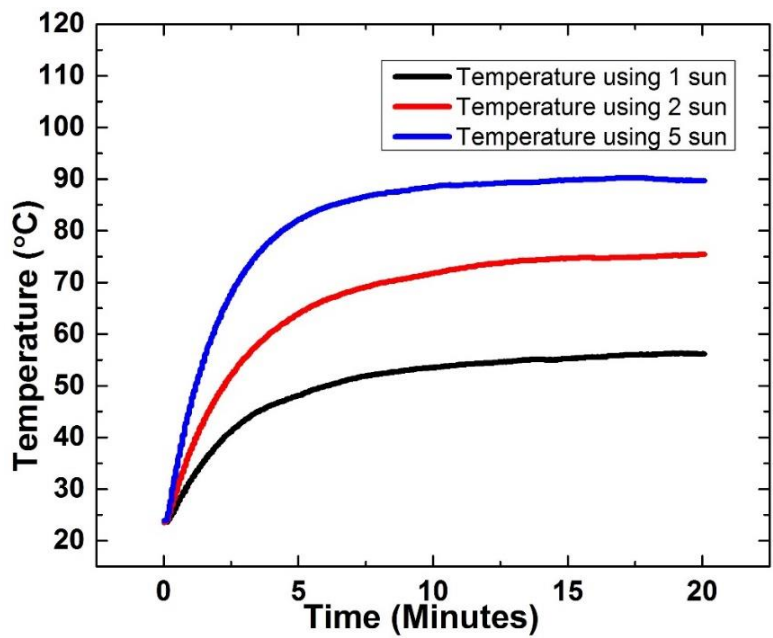

Figure 5: Solar cell temperature change using 1, 2 and 5 sun.

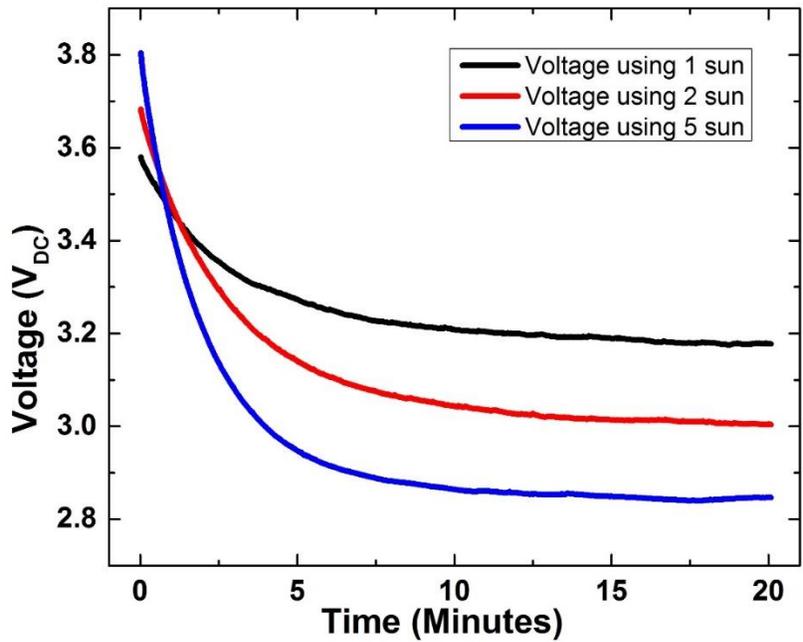

Figure 6: Solar cell open circuit voltage ($V_{OC}$) change using 1, 2 and 5 sun.

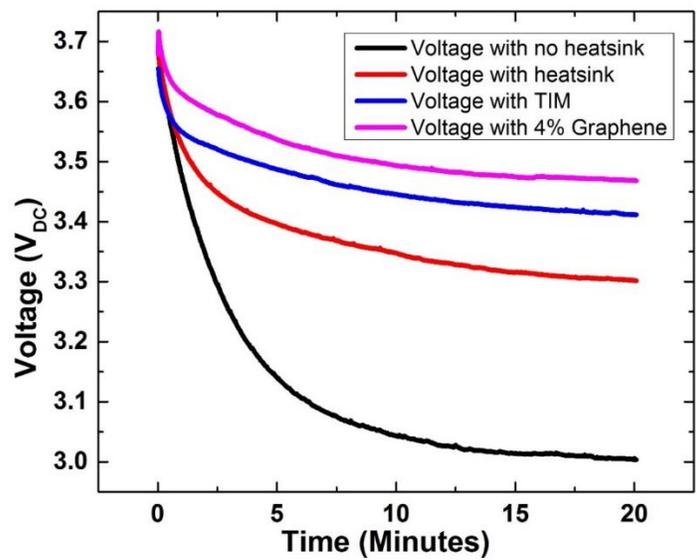

Figure 7: Open circuit voltage ($V_{OC}$) measured with and without cooling at 2 suns concentration.

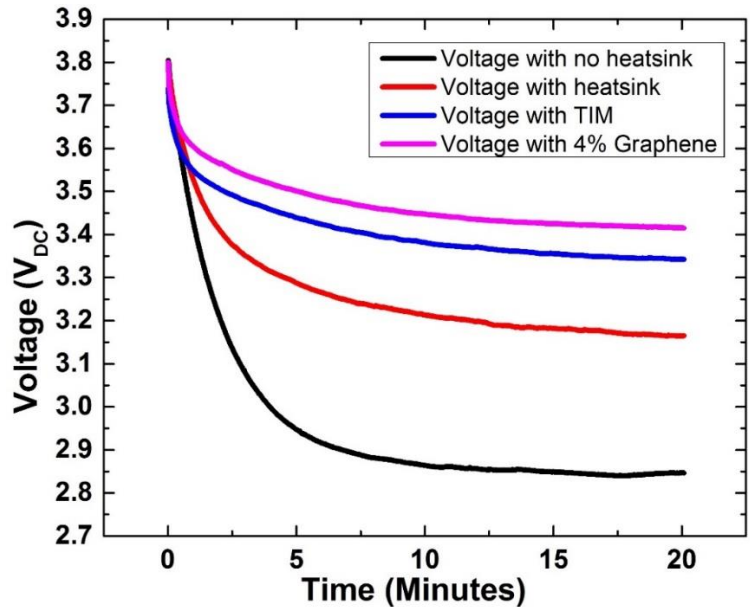

Figure 8: Open circuit voltage ($V_{OC}$) measured with and without cooling at 5 suns concentration.

## IV. CONCLUSIONS

This work presented the results of a feasibility study of the use of graphene-enhanced thermal interface materials for improving thermal management of solar cells. The increased thermal conductivity was obtained by mixing multi-layer graphene at different weight fractions with commercial TIMs. The materials were tested on solar cells using a solar simulator. The efficiency of the solar cells was improved by effectively eliminating the unwanted heat from solar cells. The introduced method can recover up to 75% of the power loss and can be commercially produced at low cost.